# Covid-19: Automatic detection from X-Ray images utilizing Transfer Learning with Convolutional Neural Networks

Ioannis D. Apostolopoulos[1], Tzani Bessiana[2]

[1] Department of Medical Physics, School of Medicine, University of Patras, 26504 Patras, Greece

[2] Computer Technology Institute, University of Patras, Patras, Greece

Correspondence: Ioannis D. Apostolopoulos, ece7216@upnet.gr

**Abstract.** In this study, a dataset of X-Ray images from patients with common pneumonia, Covid-19, and normal incidents was utilized for the automatic detection of the Coronavirus. The aim of the study is to evaluate the performance of state-of-the-art Convolutional Neural Network architectures proposed over recent years for medical image classification. Specifically, the procedure called transfer learning was adopted. With transfer learning, the detection of various abnormalities in small medical image datasets is an achievable target, often yielding remarkable results. The dataset utilized in this experiment is a collection of 1427 X-Ray images. 224 images with confirmed Covid-19, 700 images with confirmed common pneumonia, and 504 images of normal conditions are included. The data was collected from the available X-Ray images on public medical repositories. With transfer learning, an overall accuracy of 97.82% in the detection of Covid-19 is achieved.

## 1   Introduction

COVID-19 is an acute resolved disease, but it can also be deadly, with a 2% case fatality rate. Severe disease onset might result in death due to massive alveolar damage and progressive respiratory failure [1]. The early and automatic diagnosis of Covid-19 may be beneficial for countries for timely referral of the patient to quarantine, rapid intubation of serious cases in specialized hospitals, and monitoring of the spread of the disease.

Although the diagnosis has become a relatively fast process, the financial issues arising from the cost of diagnostic tests concern both states and patients, especially in countries with private health systems, or restricted access health systems due to prohibitive prices.

In March 2020, there has been an increase in publicly available x-rays from healthy cases, but also from patients suffering from Covid-19. This enables us to study the medical images and identify possible patterns that may lead to the automatic diagnosis of the disease.

The development of deep learning applications over the last five years seems to have come at the right time. Deep Learning is a combination of Machine Learning methods mainly focused on the automatic feature extraction and classification from images, while its applications are broadly met is object detection tasks, or in medical image classification tasks. Machine learning and Deep Learning have become established disciplines in applying artificial intelligence to mine, analyze, and recognize patterns from data. Reclaiming the advances of those fields to the benefit of clinical decision making and computer-aided systems is increasingly becoming nontrivial, as new data emerge [2]**.**

Deep Learning often refers to a procedure wherein deep Convolutional Neural Networks are utilized for automatic mass feature extraction, which is achieved by the process called convolution. The layers process nonlinear information [3]. Each layer involves a transformation of the data into a higher and more abstract level. The deeper we go into the network, the more complex information is learned. Higher layers of portrayal enhance parts of the information that are significant for segregation and smother unimportant attributes. Usually, deep learning refers to more deep networks than the classic machine learning ones, utilizing big data.

The purpose of this research is to evaluate the effectiveness of state-of-the-art pre-trained Convolutional Neural Networks proposed by the scientific community, regarding their expertise in the automatic diagnosis of Covid-19 from thoracic X-Rays. To achieve this, a collection of 1427 thoracic X-Ray scans is processed and utilized to train and test the CNNs. Due to the fact that the size of the samples related to Covid-19 is small (224 images), transfer learning is a preferable strategy to train the deep CNNs. This is due to the fact that the state-of-the-art CNNs are sophisticated model requiring large-scale datasets to perform accurate feature extraction and classification. With transfer learning, the retention of the knowledge extracted from one task is the key to perform an alternative task.

The results are auspicious and demonstrate the effectiveness of Deep Learning, and more specifically, transfer learning with CNNs to the automatic detection of abnormal X-Ray images related to the Covid-19 disease.

## 2 Methods

### 2.1 Dataset of the study

For the purpose of the experiments, several sources of X-Rays were accessed. Firstly, the Github Repository was analyzed for related datasets. A collection of X-Ray images from Cohen [4] was selected. Secondly, the following web sites were thoroughly examined: a) Radiological Society of North America (RSNA), b) Radiopaedia, and c) Italian Society of Medical and Interventional Radiology (SIRM). This collection is available on the Internet [5]. Thirdly, a collection of common pneumonia X-Ray scans was added to the dataset, to train the CNNs to distinguish Covid-19 from common pneumonia. This collection is available on the Internet by Kermany et al. [6].

The collected data includes 224 images with confirmed Covid-19, 700 images with confirmed common pneumonia, and 504 images of normal condition. The X-Ray images were rescaled to a size of 200x266. For the images of different pixel ration, and to avoid distortion, a 1: 1.5 ratio black background was added to achieve a perfect rescale to 200x266. The reader must note that the CNNs are capable of ignoring slight positional variance, i.e., they seek for patterns not only to a specific position of the image but also moving patterns.

## 2.2. Transfer Learning with CNNs

Transfer learning is a strategy wherein the knowledge mined by a CNN from given data is transferred to solve a different but related task, involving new data, which usually are of a smaller population to train a CNN from scratch [7].

In Deep Learning, this process involves the initial training of a CNN for a specific task (e.g., classification), utilizing large-scale datasets. The availability of data for the initial training is the most vital factor for successful training since CNN can learn to extract significant characteristics (features) of the image. Depending on the capability of the CNN to identify and extract the most outstanding image features, it is judged whether this model is suitable for transfer learning.

During the next phase, the CNN is employed to process a new set of images of another nature and to extract features, according to its knowledge in feature extraction, which was obtained during the initial training. There are two commonly used strategies to exploit the capabilities of the pre-trained CNN. The first strategy is called feature extraction via transfer learning [8] and refers to the approach wherein the pre-trained model retains both its initial architecture, and all the learned weights. Hence, the pre-trained model is used only as a feature extractor; the extracted features are inserted into a new network that performs the classification task. This method is commonly used either to circumvent computational costs coming with training a very deep network from scratch, or to retain the useful feature extractors trained during the initial stage.

The second strategy refers to a more sophisticated procedure, wherein specific modifications are applied to the pre-trained model, to achieve optimal results. Those modifications may include architecture adjustments and parameter tuning. In this way, only specific knowledge mined from the previous task is retained, while new trainable parameters are inserted into the network. The new parameters require training on a relatively large amount of data to be advantageous.

In medical tasks, the most prominent practice to perform transfer learning is exploiting the CNNs participated and stood out in the ImageNet Large Scale Visual Recognition Challenge (ILSVRC) [9], which evaluates algorithms for object detection and image classification at large scale.Data Augmentation

Data augmentation is an important technique that enhances the training set of a network and is used mainly when the training dataset contains only a few samples [16].

Geometric distortions or deformations are often utilized to either increase the number of samples for deep network training, or to balance the size of datasets. In the case of microscopical images, shift and rotation invariance, as well as robustness for deformations and grey value variations are the necessary alterations applied to each image of the training set [17].

These methods have been proven fast, reproducible and reliable. Increasing the number of the data may effectively improve the CNN's training and testing accuracy, reduce the loss, and improve the network's robustness. In the research for lung nodule detection, segmentation and classification, data augmentation techniques have been employed recently [18]. However, heavy data augmentation should be carefully considered, as this may produce unrealistic images and confuse the CNN.

### 2.3 State-of-the-art CNNs for Transfer Learning

In this section, a brief description of the CNNs employed for automatic detection is illustrated. In Table 1, the CNNs employed for the classification task and the parameters tuned for transfer learning are presented. The parameters were defined after several experiments, although the possible alternative choices are limitless and could be investigated in future research as to their contribution to the improvement of the performance. The parameter called Layer Cutoff refers to the number of untrainable layers starting from the bottom of the CNN. The rest of the layers, which are closer to the output features, are made trainable, to allow more information extraction coming from the late convolutional layers. The parameter Neural Network refers to the classifier placed at the top of the CNN to perform the classification of the extracted features. It is described by the total number of hidden layers and the total number of nodes.

**Table 1.** The CNNs of this experiment and their parameters for Transfer Learning.

| Network | Parameter | Description |
| --- | --- | --- |
| VGG19 [10] | Layer Cutoff | 18 |
| | Neural Network | 1024 nodes |
| Mobile Net [11] | Layer Cutoff | 10 |
| | Neural Network | 1000 nodes, 750 nodes |
| Inception [12] | Layer Cutoff | 249 |
| | Neural Network | 1000 |
| Xception [13] | Layer Cutoff | 120 |
| | Neural Network | 1000 nodes, 750 nodes |
| Inception ResNet v2 [12] | Layer Cutoff | 730 |
| | Neural Network | No |

All the CNNs share some common hyper-parameters. More specifically, all the convolutional layers are activated by the Rectified Linear Unit (ReLU) [14]. For the Neural Networks utilizing two hidden layers, a Dropout [15] layer is added to prevent overfitting [16]. The CNNs were compiled utilizing the optimization method called Adam [17]. The training was conducted for ten epochs, with a batch size of 64.

### 2.4 Metrics

Regarding the classification task of the CNNs, specific metrics were recorded as follows: (a) correctly identified malignant nodules (True Positives, TP), (b) incorrectly classified malignant nodules (False Negatives, FN), (c) correctly identified benign nodules (True Negatives, TN), and (d), incorrectly classified benign nodules (False Positives, FP). Please note that TP refers to the correctly predicted Covid-19 cases, FP refers to typical or pneumonia cases that were classified as Covid-19 by the CNN, TN refers to normal or pneumonia cases that were classified as non-Covid-19 cases, while the FN refers to Covid-19 cases classified as normal or as common pneumonia cases. Due to the fact that the main intention of the study is the detection of Covid-19, we measure two different accuracies. The first accuracy refers to the overall accuracy of the model in distinguishing the three classes (normal-pneumonia-Covid) and is called 3-class accuracy. The second accuracy refers to the accuracy related to Covid-19 only. That is, if an instance is typical and is classified as pneumonia by the CNN, it is still considered acceptable in terms of the presence of Covid-19. The aforementioned accuracy is called 2-class accuracy.

Based on those metrics, we compute the accuracy, sensitivity, and specificity of the model. The equations explaining the aforementioned metrics are eq. 1, 2, and 3.

$$Accuracy = (TP+TN)/(TP+TN+FP+FN) \quad (1)$$

$$Sensitivity = TP/(TP+FN) \quad (2)$$

$$Specificity = TN/(TN+FP) \quad (3)$$

## 3 Results

The training and evaluation procedure was performed with 10-fold-cross-validation. The results for each CNN are illustrated in Tables 2 and 3. In Table 2, the accuracy, sensitivity, and specificity are presented.

Table 2. Results of the CNNs used for Transfer Learning.

| Network | Accuracy 2-class (%) | Accuracy 3-class (%) | Sensitivity (%) | Specificity (%) |
|---|---|---|---|---|
| VGG19 [10] | 98.75 | 93.48 | 92.85 | 98.75 |
| Mobile Net [11] | 97.40 | 92.85 | 99.10 | 97.09 |
| Inception [12] | 86.13 | 92.85 | 12.94 | 99.70* |
| Xception [13] | 85.57 | 92.85 | 0.08 | 99.99* |
| Inception ResNet v2 [12] | 84.38 | 92.85 | 0.01 | 99.83* |

The results suggest that the VGG19 and the MobileNet achieve the best classification accuracy over the rest of the CNNs. Due to the imbalance of the dataset, all the CNNs perform seemingly well in terms of accuracy and in terms of specificity. However, as those metrics depend heavily on the number of samples representing each class, their unilateral evaluation leads to incorrect conclusions. For this reason, the combination of accuracy, sensitivity, and specificity must be the criterion for choosing the best model. The measurements denoted with an asterisk (*) in Table 2, indicate that these values are not considered acceptable in real-life problems, due to the above issue.

To further evaluate the two best models (VGG19 and MobileNet), the Confusion Matrix of each model is presented in Table 3.

**Table 3.** Confusion Matrix of the two best CNNs.

| CNN | TP | FP | TN | FN |
|---|---|---|---|---|
| VGG19 | **208** | **15** | **1189** | 16 |
| Mobile Net | 222 | 35 | 1169 | **2** |

While VGG19 achieves better accuracy, it is clear that in terms of the particular disease, the optimal results are those with the lowest number of False Negatives. A real-life interpretation of a False Negative instance would result in the mistaken assumption that the patient is not infected with what this entails for the spread of the virus and public health.

The Mobile Net outperforms VGG19 in terms of specificity and, thus, it is proven to be the most effective model for the detection of Covid-19 from X-Ray images.

## 4 Discussion

Based on the results, it is demonstrated that the transfer learning strategy with CNNs can have significant effects on the automatic detection and automatic extraction of essential features from X-ray images, related to the diagnosis of the Covid-19.

Some limitations of the particular study can be overcome in future research. In particular, a more in-depth analysis requires much more patient data, especially those suffering from Covid-19. Moreover, it is necessary to develop models capable of distinguishing Covid-19 cases from other similar viral cases, such as SARS, but also from a greater variety of common pneumonia or even physiological X-Rays. Besides, the automatic diagnosis of cases was made using only a medical image rather than a more holistic approach to the patient, based on other factors that may be offered and may behave as risk factors for the onset of the disease.

Nonetheless, the present work contributes to the possibility of a low-cost, rapid, and automatic diagnosis of the disease and can be clinically exploited in the near future. It has many advantages, the most important of which is that it does not require nursing and medical staff to contact the prospective outbreak, and this is particularly important in the spreading period of the disease.